%% file: ms.tex
\documentclass[preprint2]{aastex}
\shorttitle{NGC 1333 IRAS 4A2 Jet Rotation}
\shortauthors{Choi}

\usepackage{times}
\slugcomment{To appear in the Astrophysical Journal}

\begin{document}

\fontsize{10}{10.6}\selectfont

\title{Rotation of the NGC 1333 IRAS 4A2 Protostellar Jet}
\author{\sc Minho Choi$^1$, Miju Kang$^1$, and Ken'ichi Tatematsu$^2$}
\affil{$^1$ International Center for Astrophysics,
            Korea Astronomy and Space Science Institute,
            Daedeokdaero 776, Yuseong, Daejeon 305-348, South Korea;
            minho@kasi.re.kr.\\
       $^2$ National Astronomical Observatory of Japan,
            2-21-1 Osawa, Mitaka, Tokyo 181-8588, Japan}
\setcounter{footnote}{2}

\begin{abstract}

\fontsize{10}{10.6}\selectfont

The bipolar jet of the NGC 1333 IRAS 4A2 protostar
shows a velocity gradient in the direction perpendicular to the jet axis.
This lateral velocity gradient can be seen throughout the jet
imaged in a silicon monoxide line,
2500--8700 AU from the driving source,
and is consistent with the rotation of the accretion disk.
If this gradient is caused by the rotation of the jet around its axis,
the average specific angular momentum
is about 1.5 $\times$ 10$^{21}$ cm$^2$ s$^{-1}$.
Comparison of the kinematics between the jet and the disk suggests
that the jet-launching region on the disk has a radius of about 2 AU,
which supports the disk-wind models.
The angular momentum transported away by the jet
seems to be large enough for the protostar to accrete matter from the disk,
confirming the crucial role of jets
in the early phase of star formation process.
\end{abstract}


\section{INTRODUCTION}

In star formation processes,
from a dense core of interstellar cloud to a main-sequence star,
the specific angular momentum
has to decrease million-fold (Bodenheimer 1995),
and the loss of angular momentum is essential.
The spin-down mechanism is not well understood,
but the transportation of angular momentum by a protostellar jet
is thought to be one of the important processes (Ray et al. 2007).
The jets are thought to be driven by rotating disks
through magneto-centrifugal processes
(Blandford \& Payne 1982; Ray et al. 2007; Pudritz et al. 2007).
However, quantitative understanding of the jet-driving mechanism
has been difficult
because examples showing rotation in both disk and jet are rare. 
To understand this process quantitatively,
it is important to study the kinematics of a system
showing the rotation in both disk and jet.

NGC 1333 IRAS 4A2 is a Class 0 protostar
that belongs to a binary system
in the Perseus star-forming region at a distance of 235 pc from the Sun
(Sandell et al. 1991; Lay et al. 1995; Looney et al. 2000;
Hirota et al. 2008; Enoch et al. 2009).
IRAS 4A2 drives a prominent jet seen clearly in many molecular lines
(Blake et al. 1995; Lefloch et al. 1998).
The IRAS 4A2 bipolar jet imaged in an SiO line
is extremely well collimated and suitable for a detailed study (Choi 2005).
Comparison of the proper motion from molecular hydrogen images
with the line-of-sight velocity from the SiO spectra suggests
that the jet axis is close to the plane of the sky
with an inclination angle of $\sim$10\fdg7 (Choi 2005; Choi et al. 2006).

It was previously noticed
that the SiO jet displays a lateral velocity gradient,
which was puzzling (Choi 2005).
While rotation is an obvious explanation,
detailed interpretations were deferred
until the rotation direction of the circumstellar disk
can be determined observationally.
The disk of IRAS 4A2 is bright in the NH$_3$ lines
and elongated in the direction perpendicular to the jet axis
(Choi et al. 2007).
Recent observations showed
that the disk displays a Kerplerian-like rotation (Choi et al. 2010).
The derived mass of the central protostar is 0.08 $M_\odot$,
and the collapse age is $\sim$50,000 yr.
The sense of disk rotation is
consistent with the velocity gradient of the SiO jet,
which strongly suggests that the jet is rotating around the flow axis.

In this Letter, we present the results of
our data analysis of the NGC 1333 IRAS 4A2 SiO jet.
We describe the data and results in Section 2.
In Section 3, we discuss the rotation of the IRAS 4A2 jet
and its implications.

\section{DATA AND RESULTS}

Details of the observations and the results were presented by Choi (2005).
The NGC 1333 IRAS 4 region was observed using the Very Large Array (VLA)
in the SiO $v=0$ $J = 1 \rightarrow 0$ line.
The resulting image has a restoring beam
of FWHM = 1\farcs96 with a natural weighting.
When a higher resolution is needed for the analysis,
an image made with a robust weighting was used,
which has a beam size of 1\farcs50.
The natural weighting gives a higher signal-to-noise ratio,
while the robust weighting gives a smaller beam size.

There are several SiO emission peaks
distributed along the bipolar jet of IRAS 4A2
(see Figure 2(a) of Choi 2005). 
The downstream half of the northeastern jet (SiO outflow peaks 1--3)
is strongly disturbed by the interaction with an ambient cloud core
(Choi 2005; Baek et al. 2009).
In this Letter, we concentrate our attention
to the undisturbed part of the jet (Figure 1; SiO outflow peaks 4--12).

\input{fig1.tex}

To investigate the kinematics along the lateral direction
(i.e., perpendicular to the jet axis)
position-velocity (PV) diagrams were made
along several cuts passing through the emission peaks (Figure 2).
The reference position of each cut shown in Figure 1
is the peak position of the map integrated over the whole velocity interval.
The exact location of the reference position, however, is not important
because all the calculations in the discussion
require position differences, not the absolute position.
All the PV diagrams show
that the SiO emission is distributed along diagonal lines.
The sense of velocity gradient is consistent throughout the jet:
the northwestern side is blueshifted relative to the southeastern side.

\input{fig2.tex}

\section{DISCUSSION}

\subsection{Jet Rotation}

Since the sense of lateral velocity gradient is consistent
throughout the whole system
(the southwestern jet, the disk, and the northeastern jet),
the most natural and obvious explanation is
that the velocity gradient is caused
by the rotation around the jet/disk axis.
Alternative explanations for the lateral velocity gradients of outflow
were proposed previously (e.g., Soker 2005; Cerqueira et al. 2006),
but they do not seem to apply to the case of IRAS 4A2
(see the discussion in Section 4.3 of  Chrysostomou et al. 2008).
The linear pattern of emission in the PV diagrams (Figure 2) suggests
that either the SiO emission
comes from a cylindrical layer around the flow axis
or the SiO jet is rotating like a rigid body.
At several positions (e.g., cuts 10 and 11) the SiO emission
is stronger at both ends of the diagonal than at the middle part,
probably owing to the limb-brightening effect.
It should be noted that some emission peaks are off the diagonal lines
(e.g., the --8.1 km s$^{-1}$ peak of cut 7),
suggesting that the jet structure is
more complicated than our simple description.
There is also some ambiguity
in selecting the end points of the diagonal lines at a few cuts
(e.g., the red ends of cut 7 and cut 10).
The uncertainty introduced by this ambiguity
does not significantly affect the conclusions in Sections 3.2 and 3.3.

To understand the jet rotation quantitatively,
several physical parameters are plotted
as a function of distance from the driving source (Figure 3).
The jet radius, $R_j$, is the half-distance
between the end points of the diagonal lines shown in Figure~2.
The rotation speed, $v_\phi$, is half of the velocity difference
between the end points of the diagonal lines,
corrected for the inclination angle.
The angular speed is $\Omega = v_\phi / R_j$,
and the specific angular momentum is $L = R_j v_\phi$.
These parameters were calculated assuming
that the SiO emission at each lateral cut
(cross section perpendicular to the jet axis)
comes from a hollow ring.
Otherwise, a correction factor may be applied to each parameter.
For example, if the SiO emission comes
from a uniform-density disk with a rigid-body rotation,
the average rotation speed would be
$v_\phi$ shown in Figure 3(b) multiplied by a factor of 2/3,
and the average specific angular momentum
would need a correction factor of 1/2.
A more complete description of these parameters
would require information on the jet structure
such as rotation curve, density profile, and molecular abundance profile
as functions of radius from the jet axis,
but these details are unavailable and beyond the scope of this Letter.
Multi-species multi-transition observations in the future can be helpful.

\input{fig3.tex}

Figure 3(a) shows that $R_j$ increases almost linearly,
and extrapolating the best-fit line to the driving source
(position offset of zero) gives a small radius (100 $\pm$ 70 AU),
which suggests that the shape of the SiO jet
is similar to a cone with a finite opening angle.
Figure 4 shows a schematic description of the jet structure.
The rotation speed seems to decrease with the distance,
but the scatter is large in the downstream part of the SiO jet
(Figure 3(b)).
The angular speed also decreases with the distance:
the slope is relatively steep at small distances
and then becomes shallow at large distances (Figure 3(c)).
These behaviors of $v_\phi$ and $\Omega$ are
expected from the conservation of angular momentum
when the jet has a finite opening angle.
However, $L$ increases with the distance (Figure 3(d)),
which implies that
either the angular momentum injected into the jet may vary with time
or the correction factor varies as the jet propagates
(see the discussion in Section 3.2).
In any case, the fact that $L$ is not decreasing suggests
that the loss of angular momentum
through the interaction with the ambient medium
is not an efficient process.

\input{fig4.tex}

It is interesting to note
that $L$ of the paired data points at the same distance
from the driving source (cuts 5/7 and cuts 4/8)
agree within the uncertainties.
This physical connection between the opposite lobes of the bipolar jet
may be natural
because the jet materials at each pair
were ejected at almost the same time from the jet engine (disk).
Therefore, the physical parameters of a specific part of the jet
can be used to understand the properties of the jet engine
at the time of ejection.

\subsection{Jet Engine}

One of the interesting quantities in many models of jet driving mechanism
is the location of the jet-launching region on the disk.
Simply extrapolating $R_j$ to a small distance would not give the answer
because the intrinsic shape of the jet stream line is unknown.
Instead, the conservation laws of magneto-hydrodynamic winds can be used
(Pudritz et al. 2007).
The energy and angular momentum extracted from the disk are
in an electromagnetic form at the base (or foot) of the jet
and almost completely converted to a kinetic form
in the observed part of the jet (Anderson et al. 2003).
Therefore, $R_j$ and $v_\phi$ measured from the SiO jet
can be used to derive the angular speed at the base of the jet, $\Omega_0$,
which can be converted to the radius of outflow foot-ring, $R_f$.

The IRAS 4A2 jet was seen in a molecular hydrogen line,
and its proper motion was measured (Choi et al. 2006),
which gives the velocity of the jet in the plane of the sky
to be about 71 km s$^{-1}$.
If this value can be applied to the SiO jet,
the total specific energy and angular momentum
can be calculated at each position.
These quantities are conserved along the field line
as protostellar outflows are considered cold magneto-hydrodynamic winds
(Mestel 1968; Pudritz et al. 2007).
Then $\Omega_0$ can be obtained
by solving equation (4) of Anderson et al. (2003),
which can be converted to the foot-ring radius of the jet,
\begin{equation}
R_f = \left({{G M_*}\over{\Omega_0^2}}\right)^{1\over3},
\end{equation}
where $M_*$ is the mass of the central protostar.
Figure 5(a) shows $R_f$ at the time of ejection
for each lateral cut on the jet,
which is distributed in the range of 1.5--2.8 AU.
The average foot-ring radius is 2.0 AU.
Figure 5(a) also shows that $R_f$ is slowly increasing
with the distance or jet propagation time.
If this trend is real, it could mean
that the outflow launching region on the disk is shrinking in size.

\input{fig5.tex}

The jet, however, is not necessarily launched from a thin ring,
and the distribution of emission needs to be considered.
For example, the emission of cut 10 is mostly concentrated
at the two ends of the diagonal line in the PV diagram (Figure 2),
and the $R_f$ given in Figure 5(a) is a good representative value.
By contrast, for cut 9, strong emission
comes from the middle part of the diagonal (Figure 2).
If we take the two strongest peaks of cut 9
and repeat the same calculations,
$R_f$ can be as small as 0.6 AU.
Another implicit assumption is the Keplerian disk.
As mentioned by Choi et al. (2010),
the assumption of Keplerian rotation of accretion disk
needs to be tested with future observations.
Therefore, the uncertainties of $R_f$ and related quantities
may be larger than the statistical uncertainties shown in Figure 5.

The measured foot-ring radius of $\sim$2 AU
seems to favor the disk-wind models (Pudritz et al. 2007)
that predict jets launched from an extensive area on the disk
with a radius of a few AU.
However, jets coming from the inner edge of the disk
at a few stellar radii (Shu et al. 2000),
cannot be ruled out,
because it is possible
that there is yet another component of jet 
that is not traced by the SiO line.
Future observations of the jet rotation with atomic tracer lines
would provide a more complete answer.

\subsection{Angular Momentum Transport}

If enough mass can be loaded,
the jet can carry away the angular momentum in the disk
so that the mass accretion can continue.
The required ratio of mass ejection to mass accretion
can be approximately obtained
by comparing the specific angular momenta in the jet and disk
(Pelletier \& Pudritz 1992; Anderson et al. 2003; Pudritz et al. 2007).
The mass ejection efficiency can be calculated by
\begin{equation}
f_m = {{\dot{M}_w}\over{\dot{M}_a}}
    \approx \left({{R_f}\over{R_A}}\right)^2,
\end{equation}
where $\dot{M}_w$ is the mass outflow rate,
$\dot{M}_a$ is the mass accretion rate,
and $R_A = (L / \Omega_0)^{1/2}$ is the Alfv{\'e}n radius.
For the SiO jet of IRAS 4A2, $f_m$ is only a few percent (Figure 5(b)),
which is slightly smaller than the typical value (5--10\%)
of young stellar objects (Ray et al. 2007).
Therefore, the jet seems to be playing an essential role
in the growth of the protostar IRAS 4A2.

While rotating jets of some T Tauri stars were observed previously
(Bacciotti et al. 2002; Ray et al. 2007; Launhardt et al. 2009),
they are young stellar objects well past the main accretion phase,
and the transport of angular momentum is less critical.
Rotation of the HH 26 flow driven by a Class I source
was reported by Chrysostomou et al. (2008),
and the derived foot-ring radius is 2--4 AU.
(Also see Zapata et al. (2010) for a possible rotating jet
of a young object in an ambiguous evolutionary stage.)
It has been elusive to detect a rotating jet of a protostar
in the very early (Class 0) phase of evolution
when the removal of angular momentum is crucial.
For example, detections of lateral velocity gradients were reported
for the HH 211 jet (Lee et al. 2007),
but it was later found that the sense of gradient
can be reversed at other parts of the jet (Lee et al. 2009),
which suggests that the lateral velocity gradients in this jet
are not caused by rotation.
By contrast, the IRAS 4A2 Class~0 system presented in this Letter
shows the velocity gradient very consistently throughout the bipolar jet,
which provides reliable measures of the rotation kinematics
and allows a test of models of jet driving mechanism.

\acknowledgements

We thank Jeong-Eun Lee and Daniel T. Jaffe
for helpful discussions and suggestions.
M.C. was supported by the International Research \& Development Program
of the National Research Foundation of Korea (NRF)
funded by the Ministry of Education, Science and Technology (MEST) of Korea
(grant number: K20901001400-09B1300-03210).

\vfill
\centerline{\small\tt arXiv version}
\vspace*{-\baselineskip}

\end{document}

%% file: fig1.tex
\begin{figure}[!t]
\epsscale{1}
\plotone{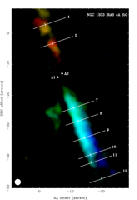}
\centerline{\scriptsize [See http://minho.kasi.re.kr/Publications.html
for the original high-quality figure.]}
\vspace{-\baselineskip}
\caption{\small\baselineskip=0.825\baselineskip
Color composite image of the NGC 1333 IRAS 4A2 bipolar jet
in the SiO $v$ = 0 $J$ = 1 $\rightarrow$ 0 line
(see Figures 2 and 3 of Choi 2005).
Blue, cyan, green, yellow, and red images show the emission
at $V_{\rm LSR}$ = --9.4, --4.7, 0.0, 14.8, and 19.6 km s$^{-1}$,
respectively, with a bandwidth of 4.7 km s$^{-1}$.
Shown in the bottom left-hand corner is the restoring beam:
FWHM = 1\farcs96.
Straight lines:
cuts for PV diagrams (Figure~2).
The PV cuts are perpendicular to the jet axis.
A short tick marks the reference position on each cut.
Solid dots:
the 3.6 cm continuum sources (Reipurth et al. 2002).}
\end{figure}

%% file: fig2.tex
\begin{figure}[!t]
\epsscale{1}
\plotone{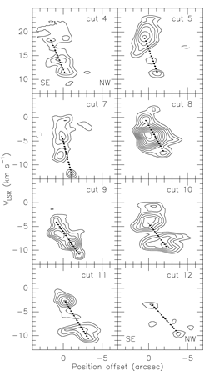}
\centerline{\scriptsize [See http://minho.kasi.re.kr/Publications.html
for the original high-quality figure.]}
\vspace{-\baselineskip}
\caption{\small\baselineskip=0.825\baselineskip
PV diagrams of the SiO line
along the cuts perpendicular to the jet axis (Figure 1).
The horizontal axis is the angular distance
from the reference position shown in Figure 1.
The PV diagrams of cuts 4, 5, 8, 10, 11, and 12
are from a map made with a natural weighting,
and the rms noise is 0.22 K.
The PV diagrams of cuts 7 and 9 are from a map made with a robust weighting,
and the rms noise is 0.40~K.
Contour levels start from 3$\sigma$ and increase by 1$\sigma$,
where $\sigma$ is the rms noise.
Dashed contours are for negative levels.
Dotted straight lines:
position-velocity gradient.}
\end{figure}

%% file: fig3.tex
\begin{figure}[!t]
\epsscale{1}
\plotone{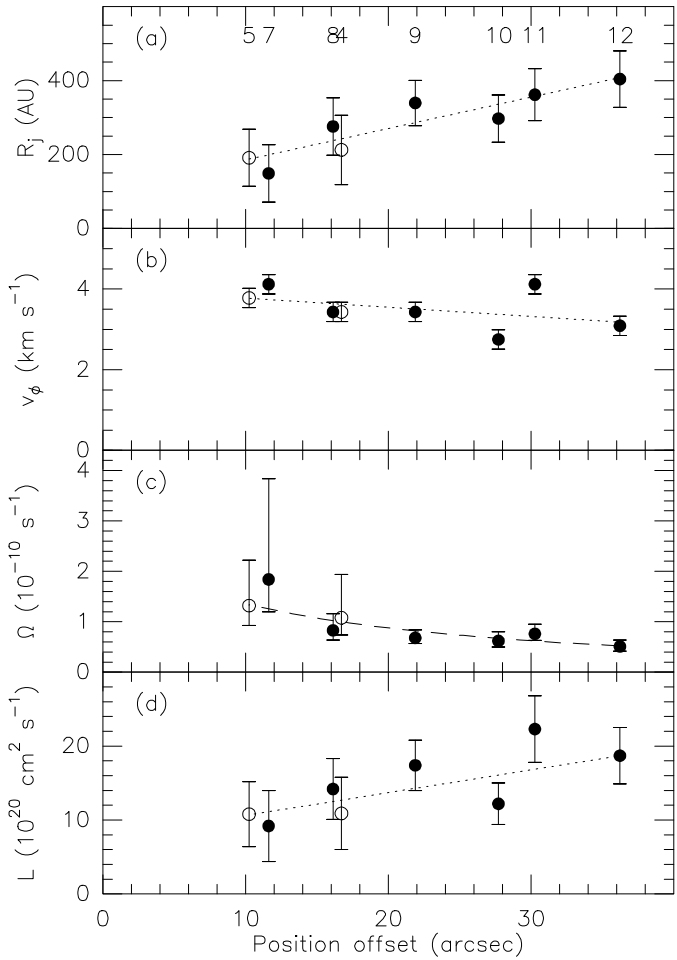}
\caption{\small\baselineskip=0.825\baselineskip
Properties of the IRAS 4A2 SiO jet.
The horizontal axis shows the angular distance from IRAS 4A2.
(a)
Radius of the jet at each lateral cut.
Names of the cuts are labeled.
Cuts 4 and 5 (open circles) are on the northeastern jet,
and cuts 7--12 (filled circles) are on the southwestern jet.
(b)
Rotation speed.
(c)
Angular speed.
(d)
Specific angular momentum.
Dotted lines:
linear fits to $R_j$, $v_\phi$, and $L$.
For $\Omega$, straight lines do not give a good description of the data.
Dashed curve:
ratio of the linear fits to $v_\phi$ and $R_j$.}
\end{figure}

%% file: fig4.tex
\begin{figure}[!t]
\epsscale{1}
\plotone{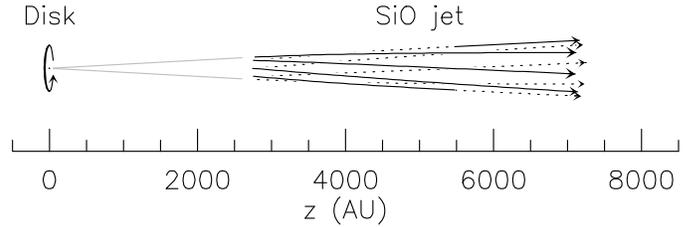}
\caption{\small\baselineskip=0.825\baselineskip
Simple schematic of the IRAS 4A2 disk-jet system.
The horizontal axis shows the distance along the jet axis
from the protostar at the center of the accretion disk.
The SiO-emitting layer of the jet is
a cone with an opening half-angle of $\sim$3\fdg1.
Loci of particles in the jet are shown
in solid curves on the near surface and in dotted curves on the far surface.
The pitch angle is $\sim$2\fdg9 in the detected part of the SiO jet.
The jet may rotate faster in the upstream part
where the SiO emission was not detected.
The jet is bipolar, one ejected to the right and the other to the left,
but only one is shown.
The counter-jet would be a mirror image of the jet shown here.
The disk has a radius of 310 AU (Choi et al. 2010).
The outflow foot-ring of 2 AU is drawn around the center of the disk,
but it is very small in this scale and appears like a point.}
\end{figure}

%% file: fig5.tex
\begin{figure}[!b]
\epsscale{1}
\plotone{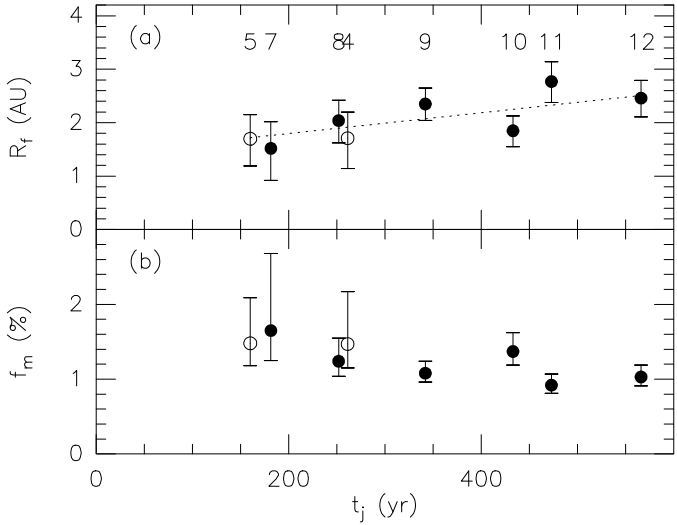}
\caption{\small\baselineskip=0.825\baselineskip
Properties of the IRAS 4A2 jet engine.
The horizontal axis shows the time of jet propagation from the driving source, assuming a constant proper motion of 0\farcs064 yr$^{-1}$
(Choi et al. 2006).
(a)
Foot-ring radius inferred from the SiO jet properties at each lateral cut.
Dotted line:
linear fit.
The slope is (1.9 $\pm$ 1.0) $\times$ 10$^{-3}$ AU yr$^{-1}$.
(b)
Mass ejection efficiency inferred from $(R_f/R_A)^2$.
The (1$\sigma$) uncertainties shown here
include statistical uncertainties only.
The uncertainty of the protostellar mass (0.08 $\pm$ 0.02 $M_\odot$)
affects these values systematically,
at a level of 10\% for $R_f$ and at a level of 20\% for $f_m$.
Note that the quantities shown here
are calculated from the end points of the diagonal lines
in the PV diagrams (Figure 2),
which is suitable when the emission is concentrated at the end points.
Otherwise, a range of values should be considered.
For example, if we take the two strong peaks of cut 9
near the middle part of the diagonal line (Figure 2),
$R_f$ can be as small as 0.6 AU, and $f_m$ can be as high as 4\%.}
\end{figure}